\documentclass[conference]{IEEEtran}
\IEEEoverridecommandlockouts

\usepackage{cite}
\usepackage{amsmath,amssymb,amsfonts}
\usepackage{graphicx}
\usepackage{textcomp}
\usepackage{xcolor}
\usepackage{booktabs}
\usepackage{multirow}
\usepackage{todonotes}
\usepackage{microtype}
\usepackage{url}
\usepackage[inline]{enumitem}
\usepackage{placeins}

\usepackage{caption}
\usepackage{subcaption}

\usepackage{listings}
\definecolor{codegreen}{rgb}{0.12,0.53,0.22}
\definecolor{codegray}{rgb}{0.40,0.40,0.40}
\definecolor{codeblue}{rgb}{0.08,0.25,0.65}
\definecolor{backcolour}{rgb}{0.97,0.97,0.98}

\lstdefinestyle{cudaStyle}{
    backgroundcolor=\color{backcolour},
    commentstyle=\color{codegreen}\itshape,
    keywordstyle=\color{codeblue}\bfseries,
    numberstyle=\tiny\color{codegray},
    stringstyle=\color{purple},
    basicstyle=\ttfamily\scriptsize,
    breakatwhitespace=false,
    breaklines=true,
    captionpos=t,
    keepspaces=true,
    numbers=left,
    numbersep=4pt,
    showspaces=false,
    showstringspaces=false,
    showtabs=false,
    tabsize=2,
    frame=lines,
    rulecolor=\color{black!30},
    xleftmargin=1.5em,
    framexleftmargin=1.5em
}
\usepackage[hidelinks]{hyperref}

\newtheorem{theorem}{Theorem}
\newtheorem{proposition}[theorem]{Proposition}

\newtheorem{corollary}[theorem]{Corollary}

\graphicspath{{arxiv_version/}{arxiv_version/figures/}{figures/}}

\def\BibTeX{{\rm B\kern-.05em{\sc i\kern-.025em b}\kern-.08em
    T\kern-.1667em\lower.7ex\hbox{E}\kern-.125emX}}

\begin{document}

\title{On the Robustness of Sparse Randomized Sketching to FP16 Rounding on GPUs}

% \author{
% \IEEEauthorblockN{Anonymous authors}
% }

\author{
\IEEEauthorblockN{Aryaman Jeendgar\textsuperscript{*}}
\IEEEauthorblockA{
Technical University of Munich\\
Germany\\
International Computer Science Institute\\
Berkeley, CA, USA
}
\and
\IEEEauthorblockN{Clément Flint}
\IEEEauthorblockA{
University of Strasbourg\\
France\\
Technical University of Munich\\
Germany
}
\and
\IEEEauthorblockN{Hartwig Anzt}
\IEEEauthorblockA{
Technical University of Munich\\
Germany\\
University of Tennessee\\
Knoxville, TN, USA
}
}

\maketitle

% Author contact information
\begingroup
\renewcommand{\thefootnote}{*}
\footnotetext{
Corresponding author: Aryaman Jeendgar
(\texttt{aryaman.jeendgar@tum.de},
\texttt{ajeendgar@icsi.berkeley.edu}).
Other authors:
Clément Flint (\texttt{cflint@unistra.fr});
Hartwig Anzt (\texttt{hartwig.anzt@tum.de},
\texttt{hanzt@icl.utk.edu}).
}
\endgroup
%%%%%%%%%%%%% NOTES to self
% \begin{enumerate}
% \color{blue}
%     \item Have to be explicit about the fact that FP16 spMM is indeed slower than FP32 spMM to explain why it doesn't show up in some of the later plots
%     \item Also discuss how a consequence of the above fact makes it non-trivial to just "do FP16 sparse sketching" naively, because even though spMM FP32 gives a solid baseline, it does not scale when doing FP16.
%     \item Check FlashSketch coherence sweep w/ FlashSketch tuned
% \end{enumerate}

\begin{abstract}
Randomized sketching replaces large matrices with smaller surrogates that preserve geometry for downstream numerical linear algebra. Sparse sketches promise faster formation than dense projections, but irregular GPU atomic updates become costly when each input row contributes to several output rows, as in SparseStack. FP16 accumulation halves sketch storage and update traffic, relieving cache pressure. How much does this rounding affect sketch geometry and downstream accuracy?

For SparseStack with four nonzeros per column, cuSPARSE's FP16 path is slower than its FP32 baseline. Our custom kernel instead pairs FP32 inputs and signed products with packed FP16 vector atomic accumulation, turning lower precision into substantial gains. On NVIDIA H200, it forms complete sketches up to $1.91\times$ faster than our FP32 baseline and up to $1.58\times$ faster than cuSPARSE FP32.

Under stated independence and range assumptions, random sketch signs cancel round-to-nearest conversion-error cross terms, giving expected squared Frobenius error no greater than ideal stochastic rounding. We assess accumulation empirically. At a sketch dimension of twice the number of input columns, the least-squares pipeline is \(1.44\)–\(1.47\times\) faster than FP32, with median paired residual increases below \(0.4\%\). Rescaling exposes range limitations and cases where stochastic conversion degrades accuracy.
\end{abstract}


\section{Introduction}

Randomized numerical linear algebra reduces the cost of large matrix computations by replacing a problem with a smaller randomized surrogate~\cite{martinsson_tropp_randnla_review,OG_sketching_sarlos,woodruff_sketching, randlapack_book}. Given $A\in\mathbb{R}^{m\times n}$, with $m\gg n$, one forms a ``sketch'' of $A$ by applying to it, a random map $S\in\mathbb{R}^{d\times m}$. The resultant matrix from the ``sketching'' operation, $SA$, retains useful geometric properties about the original~\cite{JL_lemma_og}. The reduction is useful for downstream computation only if forming and processing the sketch is cheaper than working with the original matrix.

In our work, we exclusively deal with a subclass of sketching operators, the so-called, \textit{Oblivious-Subspace-Embeddings (OSEs)}~\cite{martinsson_tropp_randnla_review, randlapack_book, Chen_online_book}. Sparse OSEs promise inexpensive sketch formation. \textbf{CountSketch} sends each input row to one random output row with a random sign~\cite{charikar_countsketch_init_paper}. Its simplicity also creates a weakness: collisions can erase directions when a few input rows carry much of the column-space energy. \textbf{SparseStack} distributes each row across $\zeta$ independent CountSketch blocks~\cite{camano_structured_sketching_tropp, chennakod_sparse_stack_theory}. These additional contributions improve robustness compared with CountSketch, but require more atomic updates on a GPU.

An appealing way to relieve this atomic-update cost is to store the sketch in a lower precision such as FP16, halving its memory footprint and the bytes associated with each update. This saving does not automatically translate into faster sketch formation. cuSPARSE provides a competitive FP32 baseline, but its FP16 sparse matrix--matrix multiplication (SpMM) path is substantially slower at small sketch dimensions for $\zeta=1,2,4$ in our measurements. Obtaining useful mixed-precision speedups therefore requires a carefully designed sparse sketching kernel.

We address this through a custom mixed-precision kernel. The input and signed products remain in FP32, while each contribution is converted with round-to-nearest (RN) and accumulated into an FP16 sketch buffer. The returned sketch is therefore stored in FP16. Packed result-discarding atomic updates process two adjacent features together, halving the sketch storage and the number of atomic instructions relative to our FP32 scatter. The kernel supports arbitrary $\zeta$; we evaluate $\zeta=1,2,4,8$, with particular interest in $\zeta=4$, a practical configuration motivated by recent work on oblivious subspace injectivity (OSI)~\cite{camano_structured_sketching_tropp}.

The numerical question is whether these savings change the information retained by the sketch. Contribution conversion and accumulation require different answers. Under stated independence and range assumptions, random signs make contribution-conversion error mean zero and cancel its cross terms in expectation, Proposition~\ref{prop:conversion}; this argument does not control the variance of a running FP16 sum. We analyze conversion error and its effect on subspace distortion Proposition~\ref{prop:distortion}, and evaluate the complete pipeline through geometry, least-squares accuracy, and controlled changes of input range.

On NVIDIA H200, our FP16 SparseStack kernel at $\zeta=4$ forms complete sketches up to $1.91\times$ faster than our FP32 scatter baseline and up to $1.58\times$ faster than cuSPARSE FP32, while closely tracking FP32 subspace distortion in the coherence sweep. It also outperforms cuSPARSE FP16, whose lower precision does not yield a speedup over vendor FP32. cuSPARSE FP32 remains faster for CountSketch. At $d=2n$, our SparseStack kernel yields $1.44$--$1.47\times$ faster complete least-squares solves than our FP32 baseline, with median paired residual increases below $0.4\%$. Larger sketches improve approximation accuracy; as FP32 factorization comes to dominate total runtime, sketch formation becomes a smaller part of the solve and accelerating it has less effect on end-to-end time. Our contributions are:
\begin{itemize}[leftmargin=*,itemsep=2pt]
\item \textbf{Fast GPU kernels for sparse sketching.} We extend the Higgins-style CountSketch scatter~\cite{higgins_GPU_sketching_paper} to SparseStack, keeping inputs and products in FP32 while accumulating and storing the output sketch in FP16 through packed result-discarding atomic updates. The FP16 kernel supports arbitrary sketch sparsity and outperforms both our FP32 scatter and cuSPARSE FP32 at $\zeta=4$. Complete sketch timings include output initialization and sketch metadata generation.
\item \textbf{Conversion-error and subspace-distortion bounds.} We prove that, under stated independence and range assumptions, random sketch signs give RN no greater expected squared Frobenius conversion error than ideal stochastic rounding. We bound the resulting change in subspace distortion and characterize conversion error under rescaling, keeping accumulation error separate.
\item \textbf{Performance, accuracy, and operating boundaries.} Paired H200 experiments across sparsities and input coherences establish complete-sketch and least-squares speedups with closely matched FP32 distortion and median paired residual increases below $0.4\%$ at $d=2n$. Comparisons with cuSPARSE distinguish the gains over our FP32 scatter from those over the vendor baseline. Common rescaling exposes FP16 range limits and the potential cost of stochastic contribution conversion.
\end{itemize}

\section{Sparse Sketches and Related Work}
\label{sec::background}

We introduce only the notation needed for the kernels and error analysis. All matrices are real; $T$ denotes transpose. Unsubscripted norms are Euclidean norms for vectors and spectral norms for matrices; $\|\cdot\|_F$ is the Frobenius norm, and $\dagger$ denotes the Moore--Penrose pseudoinverse.

\subsection{CountSketch, SparseStack, and coherence}

Let $d$ be divisible by $\zeta$. For block $t\in\{1,\ldots,\zeta\}$ and input row $i$, draw a target $h_t(i)$ uniformly from $\{1,\ldots,d/\zeta\}$ and a sign $\epsilon_{ti}\in\{-1,+1\}$ with equal probabilities. In the ideal construction, these choices are independent. With $C_t e_i=\epsilon_{ti}e_{h_t(i)}$, the normalized SparseStack matrix is
\begin{equation}
 S=\frac{1}{\sqrt{\zeta}}
 \begin{bmatrix}C_1\\ \vdots\\ C_\zeta\end{bmatrix}
 \in\mathbb{R}^{d\times m}.
 \label{eq:sparsestack}
\end{equation}
Each column of $S$ has $\zeta$ nonzeros, and $\mathbb{E}\|Sv\|^2=\|v\|^2$ for each fixed $v$: cross terms vanish under the independent signs, and the $\zeta$ copies each contribute $\|v\|^2/\zeta$. CountSketch is the case $\zeta=1$. Applying the sketch takes $\mathcal{O}(\zeta\operatorname{nnz}(A))$ arithmetic operations.

For an orthonormal basis $U\in\mathbb{R}^{m\times n}$ of $\operatorname{range}(A)$, define coherence by
\begin{equation}
 \mu(U)=\frac{m}{n}\max_i\|U_{i,:}\|^2.
 \label{eq:coherence}
\end{equation}
The row energy $\|U_{i,:}\|^2$ measures how strongly the subspace uses coordinate $i$. Since these energies sum to $n$, $1\le\mu(U)\le m/n$. A coordinate subspace $U=[I_n\;0]^T$ attains the upper limit. For this input, any CountSketch collision among the first $n$ rows makes two columns of $SU$ proportional. Its full-rank probability is therefore
\begin{equation}
 \Pr\{\operatorname{rank}(SU)=n\}
 =\prod_{j=0}^{n-1}\left(1-\frac{j}{d}\right)
 \le \exp\!\left(-\frac{n(n-1)}{2d}\right),
 \label{eq:collision}
\end{equation}
for $d\ge n$, using $1-x\le e^{-x}$. This explains why favorable results on dense random subspaces do not contradict poor worst-case CountSketch behavior.

We report subspace distortion as
\begin{equation}
 \eta(S,U)=\max\{|\sigma_{\max}(SU)-1|,
                       |1-\sigma_{\min}(SU)|\}.
 \label{eq:distortion}
\end{equation}
Thus $\eta\le\delta$ is equivalent to $(1-\delta)\|y\|\le\|SUy\|\le(1+\delta)\|y\|$ for all $y$~\cite{woodruff_sketching}. For computed sketches, we apply the formula to the returned matrix; it need not represent one exact linear operator on every input.

\paragraph{Why $\zeta=4$ and $d=2n$?}
Cama\~no et al.~\cite[Sec.~1.5.1 and Table~3]{camano_structured_sketching_tropp} identify this as an effective practical configuration, including for coordinate subspaces. Their OSI viewpoint requires $\sigma_{\min}(SU)$ bounded away from zero rather than every singular value close to one. Constant sparsity with dimension proportional to $n$ remains conjectural~\cite[Conj.~1.9]{camano_structured_sketching_tropp}; no theorem certifies these constants. We adopt their empirical recommendation and test larger dimensions and other sparsities to assess the accuracy--cost tradeoff with FP16 accumulation.

\subsection{GPU sketching}

Higgins et al.~\cite{higgins_GPU_sketching_paper} develop a high-performance CountSketch kernel for multisketching and least-squares solvers. Their scatter design is our immediate baseline; we extend it to SparseStack and study mixed-precision accumulation.

\textbf{FlashSketch}~\cite{stanford_flashsketch_paper} introduces BlockPerm-SJLT, a structured sparse sketch co-designed for GPU execution while retaining subspace-embedding guarantees. It modifies the sketch distribution for hardware efficiency, whereas we retain SparseStack and reduce its cost through FP16 accumulation and result-discarding atomic updates.

Chen et al.~\cite{chen_GPU_sketching_paper} study single- and multi-GPU sketch-and-precondition least-squares solvers using \textbf{sparse sign sketches}~\cite{sparse_sign_ose_bound}. \textbf{GraSS}~\cite{grass_paper} includes a GPU sparse projection kernel for scalable data attribution. Our focus is the performance--accuracy tradeoff of mixed-precision sparse sketching.

\subsection{Mixed precision, stochastic rounding, and dithered quantization}

Mixed precision performs dominant computations in low precision and sensitive phases in higher precision. This strategy underlies iterative refinement and Tensor Core solvers~\cite{haidar_mixed_precision_ir,carson_higham_three_precision,higham_squeezing_half_precision}, as well as HPL-AI and HPL-MxP~\cite{hpl_mxp_benchmark}, and motivates its use in bandwidth- and atomic-limited sparse sketching.

Ootomo and Yokota~\cite{ootomo_yokota_mxp_sketching} accelerate dense random projection on Tensor Cores using FP32 inputs, an FP16 projection matrix, and FP32 sketches for randomized matrix and tensor decompositions. These dense products have different hardware costs and rounding behavior from our sparse FP16 atomic accumulation.

Carson and Dau\v{z}ickait\.{e}~\cite{carson_dauzickaite_mxp_nystrom} analyze the stability and approximation accuracy of single-pass Nystr\"{o}m with lower-precision sketch formation. Their subsequent work~\cite{carson_dauzickaite_mxp_sketching_ls} analyzes sketch formation and QR in separate precisions for least-squares preconditioning, establishing conditions for an effective preconditioner and working-precision accuracy through GMRES-based iterative refinement. Our approach retains FP32 inputs, signed products, and reduced solves, restricting FP16 to contribution conversion and accumulation. We assess sketch geometry and sketch-and-solve accuracy; preconditioning, low-rank approximation, and tests of their application-specific bounds remain future work.

Deterministic RN can introduce structured, signal-dependent error. Stochastic rounding avoids scalar bias by randomly choosing adjacent values with unbiased probabilities, helping control low-precision drift~\cite{Higham_stochastic_blog,connolly_higham_stochastic}. Dither perturbs the input before RN and can decorrelate quantization error under suitable assumptions~\cite{wannamaker_dither_survey,wannamaker_dithering_thesis,dithering_stanford_pilanci}. We test both mechanisms at contribution conversion, retaining RN atomic additions. Neither provides a consistent accuracy benefit in the tested pipeline; both add random draws and conversion arithmetic that ordinary RN avoids.

% ============================================================================
\section{GPU Kernels and Precision Choices}
\label{sec::implementation}
% ============================================================================

\subsection{From CountSketch to SparseStack}
\label{subsec:sparsestack_gpu}

State-of-the-art sparse sketching on modern GPUs relies on the atomic accumulation kernel introduced by Higgins et al.~\cite{higgins_GPU_sketching_paper} for CountSketch.
In that formulation, each input row maps to a single random hash bucket ($\zeta = 1$), and each thread accumulates the scaled entry $a$ into the target row $r$ of the sketch buffer via a single \texttt{atomicAdd}.
We extend this execution model to \emph{SparseStack} in the most direct way possible: instead of one target row per input entry, each input row now contributes to $\zeta$ independent stack blocks, and we simply loop over those blocks.
Projecting the dense row-major matrix $A \in \mathbb{R}^{m \times n}$ into the sketch buffer $B \in \mathbb{R}^{d \times n}$, Listing~\ref{lst:baseline_sparsestack} details the baseline FP32 scatter implementation.

\begin{lstlisting}[float=t, language=C++, 
  caption={[FP32 SparseStack scatter streaming through input rows.]%
    FP32 SparseStack scatter streaming through input rows ($A \in \mathbb{R}^{m \times n}$, $B \in \mathbb{R}^{d \times n}$).
    Here, $\zeta$ denotes the row sparsity;
    \texttt{col} is the feature column assigned to the thread ($0 \le \texttt{col} < n$);
    \texttt{row\_idx} is the contiguous array of length $m\zeta$ storing target row indices $r \in [0, d)$;
    \texttt{val} holds the signed FP32 scales $\epsilon_{ti}/\sqrt{\zeta}$;
    and \texttt{atomicAdd} performs floating-point accumulation in global memory.}, 
  label={lst:baseline_sparsestack}]
// Thread processes column 'col', streaming across input rows i
for (int i = 0; i < m; ++i) {
    float a = A[(size_t)i * n + col];
    int base = i * zeta;
    for (int j = 0; j < zeta; ++j) {
        int r = row_idx[base + j];
        float s = val[base + j];
        float q = s * a;
        atomicAdd(&B[(size_t)r * n + col], q);
    }
}
\end{lstlisting}

The SparseStack kernel is a natural extension of the CountSketch one: when $\texttt{zeta} = 1$ the loop collapses to a single iteration and the code reduces exactly to Higgins's formulation.
It makes sure that $A$ is still being read only once by reusing the same variable $a$.
The number of \emph{atomicAdds} performed on $B$ will by construction increase as $\zeta$ increases and we rely on the GPU caches to absorb this added load.

% In practice, the CUDA code we use is close to this pseudo-code.
% One key structural difference is that we split the loop over \emph{zeta} in two, using a peeled region (up to $\texttt{j} = 7$) and a remainder loop ($\texttt{j} \geq 8$).
% The peeled region is marked with \texttt{\#pragma unroll}, so the compiler emits eight copies of the body without loop-level bookkeeping; each copy is still guarded by a runtime $\texttt{j} < \texttt{zeta}$ test, which the hardware can skip for the common small-\emph{zeta} cases.
% The remainder loop is a fallback for larger \emph{zeta} values.

\subsection{Mixed-Precision and Packed Vector Accumulation}
\label{subsec:fp16_vectorized}

While the FP32 baseline in Listing~\ref{lst:baseline_sparsestack} eliminates redundant reads of matrix $A$, memory bus throughput remains bounded by the high frequency of 32-bit atomic scatter updates to $B$.
To alleviate this bottleneck, we introduce a mixed-precision variant targeting half-precision output buffers ($B_{\text{vec}}$).
By rounding intermediate sketch projections to FP16, we halve the DRAM footprint of the sketch and exploit vectorized 32-bit memory instructions.

Listing~\ref{lst:vectorized_accumulation} illustrates this optimization.
Rather than processing a single feature column, each thread is assigned a contiguous column pair $\texttt{col0} = 2 \cdot \texttt{pair\_col}$ and $\texttt{col0} + 1$.
The thread loads both FP32 entries, evaluates their respective scaling projections, and performs scalar FP32-to-FP16 conversion via CUDA's round-to-nearest-even intrinsic \texttt{\_\_float2half}.
The resulting two 16-bit values are packed into a single 32-bit vector register (\texttt{half2}) using \texttt{\_\_halves2half2}.

\begin{lstlisting}[float=t, language=C++, 
  caption={[FP16 mixed-precision SparseStack scatter with packed updates.]%
    FP16 mixed-precision SparseStack scatter for even feature width $n$.
    The thread processes column pair $\texttt{pair\_col} \in [0, n/2)$;
    $A$ is the $m \times n$ dense FP32 input matrix;
    $B_{\text{vec}}$ is the $d \times (n/2)$ sketched output buffer reinterpreted as packed CUDA \texttt{half2}, zero-initialized;
    \texttt{\_\_float2half} executes scalar FP32-to-FP16 conversion with round-to-nearest-even (RN);
    \texttt{\_\_halves2half2} packs two FP16 scalars into a 32-bit vector;
    and \texttt{atomicAdd} denotes a single packed 32-bit dual-half atomic addition, implemented by the result-discarding helper in Section~\ref{subsec:cuda_details}.}, 
  label={lst:vectorized_accumulation}]
int col0 = 2 * pair_col;
// Thread processes column pair 'pair_col', streaming across input rows i
for (int i = 0; i < m; ++i) {
    float a0 = A[(size_t)i * n + col0];
    float a1 = A[(size_t)i * n + col0 + 1];
    int base = i * zeta;
    for (int t = 0; t < zeta; ++t) {
        int r = row_idx[base + t];
        float s = val[base + t];
        half q0 = __float2half(s * a0);
        half q1 = __float2half(s * a1);
        half2 q = __halves2half2(q0, q1);
        atomicAdd(&B_vec[(size_t)r * (n / 2) + pair_col], q);
    }
}
\end{lstlisting}

This vectorization halves the number of issued update instructions by fusing two adjacent column updates into a single packed \texttt{half2} atomic.
Atomicity is guaranteed separately for each component, not for the pair as a single transaction~\cite{nvidia_ptx_isa}: each half-precision component accumulates independently in global memory with round-to-nearest semantics.
Although packing doubles spatial locality per thread and reduces instruction issue overhead, write collisions between concurrent rows targeting the same output bucket remain an issue, just as in the FP32 baseline.
We analyze the resulting FP16 quantization noise in Section~\ref{sec:analysis}.

% \subsection{FP16 storage and packed updates}

% The mixed-precision path assigns a thread two adjacent feature columns and packs their converted contributions into a CUDA \texttt{half2}. For the even feature widths in our experiments, the packed output has row stride $n/2$, as shown in Listing~\ref{lst:vectorized_accumulation}. The packed implementations evaluated here require even feature widths.

% One packed instruction updates two FP16 components. Atomicity is guaranteed separately for each component, not for the pair as a single transaction~\cite{nvidia_ptx_isa}. Packing reduces the number of issued update instructions relative to one scalar instruction per feature; it does not remove collisions between input rows targeting the same output. Each component accumulates in FP16 with RN.

\begin{table}[t]
\centering
\caption{Precision and logical storage for $B=SA$ and the sketched least-squares solve $\min\|Ax-b\|_2$. Counts exclude metadata and a subsequent reduced solve.}
\label{tab:precision}
\small
\setlength{\tabcolsep}{4pt}
\begin{tabular}{lcc}
\toprule
Quantity & FP32 path & FP16 path\\
\midrule
Input $A$ storage & $4mn$ bytes & $4mn$ bytes\\
Scale/product arithmetic & FP32 & FP32\\
Contribution to atomic & FP32 & FP16\\
Accumulation and $B$ & FP32 & FP16\\
Output $B$ storage & $4dn$ bytes & $2dn$ bytes\\
Update instructions (even $n$) & $\zeta mn$ & $\zeta mn/2$\\
Right-hand-side sketch $Sb$ & FP32 & FP32\\
Reduced QR and solve & FP32 & FP32\\
\bottomrule
\end{tabular}
\end{table}

The routing and scale metadata are stored uncompressed: each of the $m\zeta$ entries carries an \texttt{int32} target row and a signed FP32 scale $\epsilon_{ti}/\sqrt{\zeta}$. Only the sign varies per entry (the magnitude $1/\sqrt{\zeta}$ is determined by $\zeta$), so the scale could in principle be encoded in a single bit; we nonetheless store the full FP32 value, since the metadata is small ($m\zeta$ entries) and generated once per sketch, and applying the stored value directly avoids re-deriving the magnitude inside the scatter loop.
Table~\ref{tab:precision} makes the FP16 saving explicit: the dense input $A$ is still read in FP32 ($4mn$ bytes), but the sketch buffer shrinks from $4dn$ to $2dn$ bytes and each atomic contribution writes two bytes instead of four. Counting one read and one write per contribution gives the logical traffic
\begin{equation}
T_{\rm logical}(b)=4mn+2b\zeta mn,
\qquad b\in\{4,2\}\text{ bytes},
\label{eq:traffic}
\end{equation}
i.e.\ $4\zeta mn$ atomic bytes in FP16 versus $8\zeta mn$ in FP32 at the same $\zeta$; packing two components into one \texttt{half2} then halves the number of issued atomic instructions to $\zeta mn/2$ without changing these byte counts.
Because the $4mn$-byte FP32 input read is common to both paths, the total ratio at $\zeta=4$ is $36/20=1.8$ rather than $2$.
Overall, the FP16 path halves both the output storage and the number of update instructions relative to FP32, at the price of the quantization noise quantified in Section~\ref{sec:analysis}; the kernel structure and the surrounding pipeline are otherwise unchanged.

\subsection{CUDA implementation details}
\label{subsec:cuda_details}

In practice, the CUDA code we use is close to the pseudo-code of Listings~\ref{lst:baseline_sparsestack} and~\ref{lst:vectorized_accumulation}.
One structural difference is that we split the loop over \emph{zeta} in two, using a peeled region (up to $\texttt{j} = 7$) and a remainder loop ($\texttt{j} \geq 8$).
The peeled region is marked with \texttt{\#pragma unroll}, so the compiler emits eight copies of the body without loop-level bookkeeping; each copy is still guarded by a runtime $\texttt{j} < \texttt{zeta}$ test, which the hardware can skip for the common small-\emph{zeta} cases.
The remainder loop is a fallback for larger \emph{zeta} values; this keeps a single kernel valid for arbitrary sparsity while bounding the unrolled code size.
This structure is shared by the FP32 baseline and the packed FP16 kernel.

Adjacent lanes of a warp are assigned to consecutive columns (or column pairs), so the reads from $A$ are coalesced across the warp.
We launch thread blocks of size $32 \times 16$ (i.e., $32 \times 16$ values in FP32, $64 \times 16$ values in FP16), with input rows indexed along the first grid dimension and columns (or column pairs) along the second; each thread streams over input rows in its assigned column, as in the listings above.
The routing and sign metadata (\texttt{row\_idx}, \texttt{val}) is generated once per sketch by a coupled Philox kernel on the GPU and stored as contiguous \texttt{int32} and FP32 arrays of length $m\zeta$; all kernels in this work consume this same metadata.

Listing~\ref{lst:vectorized_accumulation} uses \texttt{atomicAdd} to denote packed addition of both FP16 components at an address of the sketch buffer. The standard CUDA library overload on \texttt{half2} would compile on Hopper (compute capability 9.0) to the SASS instruction \texttt{ATOM.E.ADD.F16x2.RN}, a global-memory atomic with a destination operand and a predicate.
With an unused C++ return value, \texttt{ptxas} routes the destination to the zero register (\texttt{RZ}), a write-only sink, but the instruction still carries a destination slot and a predicate, \textit{i.e.}, a return path our scatter never uses.

Our FP16 kernel implements this packed addition with an explicit global-memory atomic whose result is discarded, emitted as a one-line inline-PTX helper:

\begin{lstlisting}[language=C++,caption={Result-discarding packed atomic.}]
asm volatile("atom.global.add.noftz.f16x2 %0, [%1], %2;"
             : "=r"(old) : "l"(ptr), "r"(packed) : "memory");
\end{lstlisting}

At the PTX level the only difference is the \texttt{.global} qualifier (\texttt{atom.add} $\rightarrow$ \texttt{atom.global.add}).
At the SASS level, this qualifier lets \texttt{ptxas} emit the returnless reduction \texttt{REDG.E.ADD.F16x2} in place of \texttt{ATOM}: both perform the same nearest-even FP16 addition at the memory location, but the reduction has no destination operand (its encoding carries none), so no result is written back to the SM.
The \texttt{G} is Hopper's descriptor-addressed form of the reduction family (the plain \texttt{RED} of earlier architectures), targeting memory through the 256-bit descriptors of compute capability 9.0.

The operation is fire-and-forget: the kernel issues one atomic per contribution and reads none of the results, so no return-path traffic reaches the SM.
This fire-and-forget behavior avoids returning the previous value to the SM.

Both instruction forms carry the same FP16 subnormal policy. \texttt{.noftz} is a PTX-level qualifier: the CUDA library's \texttt{atomicAdd} overload emits \texttt{atom.add.noftz.f16x2}, and we keep the identical qualifier in our inline assembly, so the arithmetic matches exactly. \texttt{.noftz} disables flush-to-zero and preserves FP16 subnormals; this matters for sketching, because contributions are scaled by $1/\sqrt{\zeta}$ and small input entries can fall into the subnormal range, where flushing would silently drop them from the accumulated sum. The SASS mnemonics \texttt{ATOM.E.ADD.F16x2.RN} and \texttt{REDG.E.ADD.F16x2.RN} do not spell out this policy, but both instructions inherit the same semantics from their PTX, so the subnormal behavior is identical across the two instruction forms.

The FP16 kernel shares the runtime \emph{zeta}, $32 \times 16$ thread blocks, and FP32 products with the FP32 scatter; contribution conversion, accumulation, and output storage use FP16.

\subsection{Conversion rules}

In this work we consider three rounding methods for the FP32-to-FP16 conversion of each contribution:
\begin{itemize}
    \item \emph{Round-to-Nearest (RN)}:~the native nearest-even conversion used throughout Section~\ref{subsec:fp16_vectorized} (intrinsic \texttt{\_\_float2half}); it is the conversion of the kernel presented above.
    \item \emph{Dither}:~the contribution is perturbed before RN conversion using the historical relative-noise construction
    \begin{equation}
      Q_{\rm dith}(q)=\operatorname{RN}_{16}\bigl(q+(u_0-\tfrac12)cq\bigr),
      \label{eq:dither}
    \end{equation}
    with a 24-bit variate $u_0\in[0,1)$ from an \texttt{xorshift} generator and the FP32 coefficient $c$ written as \texttt{0.00097656f}, approximately $2^{-10}$; the perturbation is evaluated in FP32.
    \item \emph{Stochastic Rounding (SR)}:~the contribution is rounded to one of its neighboring FP16 values $l\le q\le v$, selecting $v$ with probability approximately $(q-l)/(v-l)$ and $l$ otherwise, and returned unchanged when $q$ is representable. The ideal rule has expectation $q$.
\end{itemize}

All three paths feed the same packed atomic accumulation; they differ only in how the contribution is converted. In our kernels, dither adds one \texttt{xorshift} draw and the perturbation arithmetic to the update loop. Its relative width $c|q|$ is not uniformly one FP16 spacing, in particular at exponent boundaries and in the subnormal range, so this construction does not inherit ideal-SR unbiasedness or its variance formula.

Stochastic rounding adds the computation of the neighboring FP16 endpoints and a random decision to the update loop. Because our finite random-bit implementation need not realize all such probabilities exactly, we call the experimental path stochastic conversion and reserve \emph{ideal SR} for the mathematical model.

As we will see, RN has no greater expected squared contribution-conversion error than ideal SR under the assumptions of Proposition~\ref{prop:conversion}; this comparison does not extend to accumulation.
Mixed-precision work routinely reaches for dither and stochastic rounding, and they often help in other settings, but in our case neither provides a consistent accuracy benefit in the range experiments (Section~\ref{subsec:range}). We therefore retain them as diagnostic controls alongside the simpler RN path, which avoids the extra random draws and conversion arithmetic described above.

\section{Error Analysis and Operating Conditions}
In this section, we formalize our rounding pipeline and derive straightforward results for contribution conversion under stated assumptions. We prove a sign-averaged comparison between RN and ideal SR and derive conservative bounds on conversion error and its effect on subspace distortion. These bounds do not fully explain complete-kernel accuracy and are looser than the FP16-FP32 agreement we observe in both distortion and least-squares accuracy: FP16 accumulation remains outside the conversion guarantee and is assessed empirically, Section~\ref{sec::empirics}

\label{sec:analysis}

\subsection{Error model}

Take the stored FP32 input $A$ as exact and let $B=SA$ be its ideal sketch. Let $B_q$ sum the actual signed FP32 products exactly, $B_c$ sum their individually converted FP16 values exactly, and $D$ be the returned FP16 sketch. Adding and subtracting these intermediate quantities gives
\begin{equation}
 E:=D-B=
 \underbrace{B_q-B}_{P\text{: scale/product}}
 +\underbrace{B_c-B_q}_{\Delta_c\text{: conversion}}
 +\underbrace{D-B_c}_{\Delta_a\text{: accumulation}}.
 \label{eq:error-split}
\end{equation}
The identity and $\|E\|\le\|P\|+\|\Delta_c\|+\|\Delta_a\|$ require no independence. Figure~\ref{fig:error-pipeline} gives an idealized representation of our rounding pipeline in terms of these introduced quantities.

\begin{figure}[!ht]
\centering
\begin{tikzpicture}[x=1cm,y=1cm,>=stealth,
  stage/.style={draw=black!55,rounded corners=1pt,fill=black!3,
    minimum height=0.65cm,text width=1.40cm,align=center,
    inner sep=2pt,font=\footnotesize},
  check/.style={font=\footnotesize,inner sep=2pt},
  note/.style={font=\scriptsize,align=center,fill=white,inner sep=1pt}]
 \node[stage,text width=1.0cm] (a) at (0,0) {FP32\\input $A$};
 \node[stage] (q) at (2.0,0) {FP32\\products};
 \node[stage] (v) at (4.25,0) {FP16\\updates};
 \node[stage] (acc) at (6.50,0) {FP16 RN\\atomics};
 \draw[->] (a) -- node[note,above,yshift=10pt] {scale/sign} (q);
 \draw[->] (q) -- node[note,above] {$Q$} (v);
 \draw[->] (v) -- (acc);
 \node[check] (b) at (0,-1.15) {$B=SA$};
 \node[check] (bq) at (2.0,-1.15) {$B_q$};
 \node[check] (bc) at (4.25,-1.15) {$B_c$};
 \node[check] (d) at (6.50,-1.15) {$D$\;\textnormal{(FP16)}};
 \draw[->,dashed] (a) -- node[note] {ideal $S$} (b);
 \draw[->,dashed] (q) -- node[note] {exact sum} (bq);
 \draw[->,dashed] (v) -- node[note] {exact sum} (bc);
 \draw[->] (acc) -- (d);
 \draw[->,black!65] (b) -- node[note,below] {$P$} (bq);
 \draw[->,black!65] (bq) -- node[note,below] {$\Delta_c$} (bc);
 \draw[->,black!65] (bc) -- node[note,below] {$\Delta_a$} (d);
\end{tikzpicture}
\caption{FP16 sketching pipeline. Dashed branches define analytical references; the kernel returns only $D$. Bottom arrows label the differences in~\eqref{eq:error-split}, not additional kernel operations. All sums use the same bucket routing.}
\label{fig:error-pipeline}
\end{figure}

For input row $i$, stack block $t$, and feature $j$, write the actual product as $\epsilon_{it}q_{itj}$, where $\epsilon_{it}\in\{-1,1\}$ is the sketch sign and $q_{itj}$ is the product with that sign removed. Condition on the sign-removed products $q_{itj}$ and their target buckets, and assume the remaining signs are independent and equiprobable. This requires sign-equivariant product evaluation; $q_{itj}$ may itself be negative. The term $P$ retains scale and FP32 multiplication errors. At $\zeta=4$, multiplication by the ideal scale $1/2$ is exact whenever representable, provided that scale is stored exactly.

We assume gradual underflow, no overflow, and finite FP16 neighbors bracketing every product. Conversion $Q$ is either nearest-even RN or ideal SR, which selects the neighbors with exact unbiased probabilities and independent draws, also independent of the signs. All policies subsequently use RN atomic addition. Our finite-bit SR and relative dither implementations detailed in Section~\ref{sec::implementation} provide empirical emulations of randomized-conversion mechanisms, without inheriting ideal-SR guarantees from this model. The observed stochastic-conversion range trends (Section~\ref{subsec:range}) are consistent with the variance mechanism predicted by ideal SR.

\subsection{Conversion: cancellation from sketch signs}

\begin{proposition}[Sign-averaged conversion error]
\label{prop:conversion}
Let $e_{itj}=Q(q_{itj})-q_{itj}$. Under the stated assumptions, RN and ideal SR satisfy $\mathbb E\Delta_c=0$ and
\begin{align}
 \mathbb E\|\Delta_c\|_F^2
 &=\sum_{itj}\mathbb E e_{itj}^2,\label{eq:conversion-identity}\\
 \mathbb E\|\Delta_c\|_F^2
 &\le u^2\!\sum_{|q_{itj}|\ge a}q_{itj}^2
       +\frac{h^2}{4}N_{\rm sub}=:V,\label{eq:conversion-bound}
\end{align}
where $u=2^{-11}$, $a=2^{-14}$, $h=2^{-24}$, and $N_{\rm sub}$ counts products with $0<|q_{itj}|<a$. For the same products and routing, RN has no larger expected squared Frobenius conversion error than ideal SR.
\end{proposition}
\begin{IEEEproof}
RN is odd: $Q(-q)=-Q(q)$. Its signed error is therefore $\epsilon_{it}e_{itj}$, with mean zero. Squaring one bucket sum gives diagonal terms $e_{itj}^2$ and cross terms containing $\epsilon_{it}\epsilon_{kt}$ for $i\ne k$. Independence makes their expectations zero. Summing over output entries counts each contribution once, proving~\eqref{eq:conversion-identity}; signs shared across features do not affect this entry-wise calculation.

Ideal-SR errors are independent and mean zero conditional on the signs, so their cross terms also vanish. Reflection changes the scalar error sign in distribution, preserving its second moment. For $q=l+pw$ between adjacent FP16 values, with spacing $w$ and $0<p<1$, the errors are $(1-p)w$ with probability $p$, and $-pw$ otherwise. Thus
\[
 \mathbb E e_{\rm SR}^2=p(1-p)w^2,
 \qquad e_{\rm RN}^2=\min(p,1-p)^2w^2.
\]
The latter is no larger than the former, and both are at most $w^2/4$. For $|q|\ge a$, $w\le2u|q|$; below $a$, $w=h$. Representable values contribute zero. Summation proves the bound and the comparison.
\end{IEEEproof}

The RN expectation averages sketch signs; ideal SR also averages its rounding draws. Repeating one fixed RN sketch is not this expectation. With exact normalization and products, $\sum_{itj}q_{itj}^2=\|A\|_F^2$: each entry appears $\zeta$ times with squared scale $1/\zeta$. The normal-range bound is therefore $u^2\|A\|_F^2$, with no extra factor of $\zeta$.

The common bound does not make the policies equivalent. RN minimizes scalar squared conversion error, and the signs cancel its cross terms. We therefore conclude that unbiased scalar conversion as guaranteed by SR is therefore unnecessary for achieving an unbiased conversion error in our model.

\subsection{Accumulation: the cancellation argument stops}

For one component of a packed \texttt{half2} atomic sum, let $v_1,\ldots,v_L$ be the converted updates in their realized order. Set $z_0=0$, $z_k=\operatorname{RN}_{16}(z_{k-1}+v_k)$, and $a_k=z_k-(z_{k-1}+v_k)$. Then
\begin{align*}
 z_L-\sum_{k=1}^L v_k&=\sum_{k=1}^L a_k,\\
 \mathbb E\Bigl(\sum_k a_k\Bigr)^2
 &=\sum_k\mathbb E a_k^2+2\sum_{k<\ell}\mathbb E[a_ka_\ell].
\end{align*}
Each $a_k$ depends on the preceding rounded sum, so the cross terms need not vanish. For example, at $z=1$, adding $+2^{-11}$ rounds back to one, whereas adding $-2^{-11}$ is exact: equally likely updates have conditional mean addition error $-2^{-12}$. This does not imply a nonzero overall mean. For RN conversion, finite sums, and a fixed sign-independent order, reversing all symmetric signs reverses the final error; zero mean still leaves its variance uncontrolled by Proposition~\ref{prop:conversion}.

A conservative serial bound for finite sums is
\[
 \left|z_L-\sum_{k=1}^L v_k\right|
 \le\frac{Lu}{1-Lu}\sum_{k=1}^L|v_k|,\qquad Lu<1.
\]
Each addition has relative error at most $u$; a subnormal exact sum is a multiple of $h$ and already representable. Expanding the rounding factors gives this bound. However, the mean bucket depth $m\zeta/d$ is already 4096 at a representative operating regime of our experiments in Section~\ref{sec::empirics}, $m=2^{21}$, $d=2048$, $\zeta=4$, so its product with $u$ is two. This estimate does not apply at those depths, and doesn't supply an explanation of our observed accuracy in both the sense of distortion and least squares error. Parallel atomic updates may also realize different orders across executions. We therefore retain $\Delta_a$ explicitly in further results and assess complete-kernel behavior empirically.

\subsection{Effect on subspace distortion}

\begin{proposition}[Distortion perturbation]
\label{prop:distortion}
For $X,Y\in\mathbb R^{d\times n}$, $d\ge n$, extend~\eqref{eq:distortion} by $\eta(X)=\sup_{\|v\|=1}|\|Xv\|-1|$. Then
\begin{equation}
 |\eta(X)-\eta(Y)|\le\|X-Y\|_2\le\|X-Y\|_F.
 \label{eq:distortion-perturbation}
\end{equation}
In particular, for sketches of an orthonormal basis $U$, Proposition~\ref{prop:conversion} implies
\begin{equation}
 \mathbb E|\eta(B_c)-\eta(B_q)|^2\le V.
 \label{eq:conversion-distortion}
\end{equation}
With exact normalization and products, $V\le u^2n+h^2N_{\rm sub}/4$.
\end{proposition}
\begin{IEEEproof}
For each unit $v$, two reverse triangle inequalities give
\[
 \bigl|\,|\|Xv\|-1|-|\|Yv\|-1|\,\bigr|
 \le\|(X-Y)v\|\le\|X-Y\|_2.
\]
Taking suprema proves~\eqref{eq:distortion-perturbation}. Apply it to $B_c-B_q=\Delta_c$, square, and take expectations to obtain~\eqref{eq:conversion-distortion}. Finally, $\|U\|_F^2=n$ gives the stated bound on $V$.
\end{IEEEproof}

These are conservative bounds, not predictions of the measured distortion differences. Even with all products normal, $u\sqrt n=1/64$ at $n=1024$ is only an upper bound on RMS conversion-induced distortion change. The much smaller observed differences include accumulation: with the same map and FP32 products,
\[
 D_{16}-D_{32}=\Delta_c+\Delta_{a,16}-\Delta_{a,32}.
\]
Thus the proposition bounds the geometric effect of conversion, but does not certify complete FP16--FP32 agreement we observe in Section~\ref{subsec:coherence}. Conversely, similar scalar distortions do not bound the matrix difference.

\subsection{Conversion under rescaling}

The following specialization of Proposition~\ref{prop:conversion}'s scalar calculation explains the range experiment. Compare errors in original units: for exact power-of-two scaling, set $e_k=2^{-k}Q(2^kq)-q$.

\begin{corollary}[Conversion below the first FP16 grid point]
\label{cor:range-scaling}
If $0<2^k|q|<h/2$, RN gives $e_k=-q$. For ideal SR with $0<2^k|q|<h$,
\begin{equation}
 \mathbb E e_k=0,\qquad
 \mathbb E e_k^2=h|q|2^{-k}-q^2.
 \label{eq:scaled-sr-variance}
\end{equation}
\end{corollary}
\begin{IEEEproof}
RN returns zero below $h/2$. In the SR formula, substitute spacing $w=h$ and nonzero-endpoint probability $p=2^k|q|/h$, then multiply the variance by $2^{-2k}$ to undo the scale. This gives~\eqref{eq:scaled-sr-variance}.
\end{IEEEproof}

Erased RN contributions have constant error in original units. For $2^k|q|\ll h$, ideal-SR RMS error instead grows as $2^{-k/2}$. Equation~\eqref{eq:conversion-identity} transfers this law to the RMS Frobenius conversion error of a fixed set of tiny contributions. Section~\ref{subsec:range} examines its empirical relevance; the corollary includes neither atomic accumulation nor finite-bit SR.

\section{Empirical Evaluation}
\label{sec::empirics}
% Retain the expanded Higgins and range explanations; recover space primarily from introductory framing and float layout.

% Compact spacing applies only from the empirical section onward.
\setlength{\dblfloatsep}{4pt plus 1pt minus 1pt}
\setlength{\dbltextfloatsep}{6pt plus 1pt minus 1pt}
\setlength{\textfloatsep}{6pt plus 1pt minus 1pt}
\captionsetup[figure]{skip=3pt}

We evaluate whether the kernel speedups preserve sketch geometry and carry through to least-squares solves. We then rescale the inputs without changing the mathematical problem, to expose where FP16 representation limits alter the result.

\begin{figure*}[!t]
\centering
\includegraphics[width=0.95\textwidth,trim=0pt 5pt 0pt 2pt,clip]{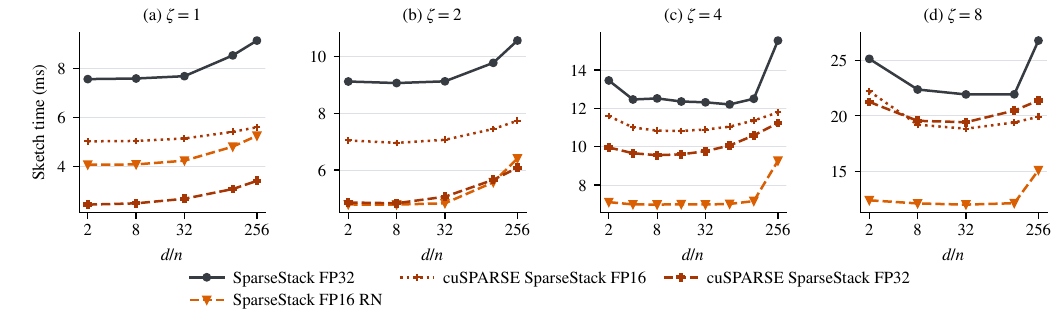}
\caption{Complete sketch time on dense $m=2^{21}$, $n=1024$ inputs at $\zeta=1,2,4,8$. Colors retain their meaning across panels. Both vendor storage formats include GPU CSR construction. Markers show measured dimensions; lines and bands give seed medians and observed ranges.}
\label{fig:h200-tuning}
\end{figure*}

\subsection{Setup and reporting}
\label{subsec:setup}

The main experiments use one H200 NVL: driver 610.57.04, CUDA compiler 13.3.33, PyTorch 2.13.0/CUDA 13.0, and cuSPARSE 12.6.3. Custom kernels target SM90 with TF32 disabled. We compare our FP32 scatter, SparseStack FP16 RN kernel (\emph{FP16 SS}), and cuSPARSE, with CountSketch and FlashSketch as additional controls. A separate CountSketch comparison uses H100 NVL with CUDA~12.4.1/cuSPARSE~12.3.1.

The H200 grids cover matrix size, sketch dimension, sparsity, coherence, and input scale. Unless specified otherwise, $n=1024$, $\zeta=4$, and matrix seed 123 is fixed. We use ten independent sketch seeds, with 11 timing repetitions in the sketch grid and five elsewhere; the H100 control uses 100 paired seeds. FlashSketch tuning uses 24 parameter combinations and three training seeds, with disjoint evaluation seeds. Repetition medians precede seed summaries, and ratios and differences pair seeds before aggregation. Bands are observed seed ranges, not confidence intervals.

Complete sketch CUDA-event times include initialization, map generation, and application. cuSPARSE builds sorted CSR on the GPU, using FP32 ALG2 or FP16 DEFAULT; the latter downcasts inputs. Complete LS wall time includes shared-map $SA,Sb$, layout/promotion, FP32 QR, projection, and solve. Input generation, compilation, references, and diagnostics are excluded. FP64 evaluates residuals; unsketched FP64 QR supplies the stored-system solution reference.

The main FP16 SS results use the current kernel. For the SR and dither controls, we take the median over three conversion draws within each sketch seed before summarizing the ten sketch seeds.

\begin{figure}[!t]
\centering
\includegraphics[width=0.84\columnwidth,trim=0pt 5pt 0pt 4pt,clip]{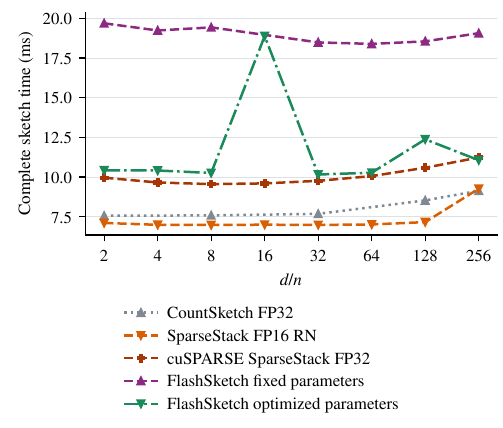}
\caption{Complete sketch time at $m=2^{21}$, $n=1024$. SparseStack uses $\zeta=4$; CountSketch uses one contribution and its five-point grid. The fixed and optimized parameters share the same recorded measurement at $d/n=16$.}
\label{fig:h200-external}
\end{figure}

\subsection{Sparsity, precision, and complete sketch cost}
\label{subsec:kernel-performance}

The timing grid uses $m\in\{2^{19},2^{20},2^{21}\}$, dense inputs with singular values decreasing linearly from 1000 to 1, and coordinate inputs $[I_n\;0]^T$. Figure~\ref{fig:h200-tuning} shows the largest dense case. We sample $d/n=2,4,8,16,32,64,128,256$ at $\zeta=4$ and $2,8,32,128,256$ elsewhere, extending well beyond the compact sketches used in our primary least-squares experiments.

At $\zeta=4$, FP16 SS accelerates complete sketch formation by up to $1.91\times$ over our FP32 scatter and $1.58\times$ over cuSPARSE FP32. The largest dense input at our primary operating point $d=2n$ already achieves a speedup close to $1.9\times$ over FP32 scatter. Sketch time is relatively flat over compact dimensions and rises at the largest tested ratios, where the sketch itself becomes much larger. We test the largest sketch ratios ($d/n>32$) to the end of producing a more complete performance story rather than as a recommendation for practical usage, where we still advocate for the suggestion of Camaño et al.~\cite{camano_structured_sketching_tropp} i.e. $d/n=2$ with $\zeta>=4$

cuSPARSE FP32 remains faster for CountSketch ($\zeta=1$), whereas FP16 SS is faster in every tested $\zeta=4/8$ case. cuSPARSE's FP16 path does not provide the same performance benefit: at $m=2^{21}$, $d=2n$, and $\zeta=4$, it takes 11.60\,ms versus 9.96\,ms for cuSPARSE FP32 and 7.11\,ms for FP16 SS. This path also pays to downcast the FP32 input. The measured regression, particularly at small sketch dimensions, shows why obtaining faster FP16 sparse sketching requires more than simply selecting a lower-precision SpMM format.

\begin{figure}[!t]
\centering
\includegraphics[width=0.84\columnwidth,trim=0pt 4pt 0pt 0.5pt,clip]{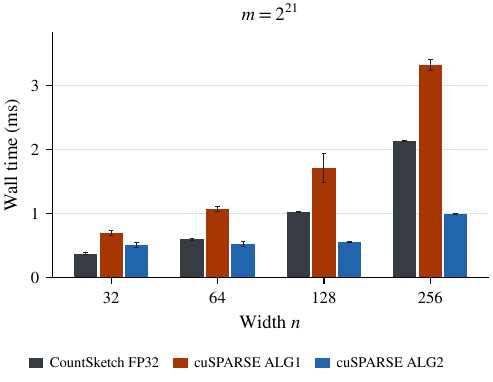}
\caption{H100 NVL CountSketch control: mean complete wall time, including sparse construction for cuSPARSE CSR ALG1 and ALG2, at $m=2^{21}$ and $d=2n^2$. Bars show standard deviations across paired seeds. This is a representative slice of the full control.}
\label{fig:h100-higgins-control}
\end{figure}

\paragraph{Reconciling the cuSPARSE comparison}
Higgins et al.~\cite{higgins_GPU_sketching_paper} report a CountSketch scatter kernel substantially faster than cuSPARSE SpMM. Because this comparison motivates the FP32 kernel on which we build, we additionally test their full set of eleven matrix shapes and $d=2n^2$ in a separate H100 control using the same CUDA~12.4 environment used in their work. Figure~\ref{fig:h100-higgins-control} shows the $m=2^{21}$ slice. Scatter beats CSR ALG1 throughout, but CSR ALG2 reverses the ordering at every tested $n\ge64$, including sparse construction. At $n=256$ in the plotted slice, ALG2 is more than twice as fast as scatter.

The reported cuSPARSE disadvantage is therefore not reproduced by our optimized baseline, and the discrepancy changes the CountSketch performance story materially. Higgins et al. do not specify their SpMM algorithm or sparse storage format, so these results do not identify the source of the discrepancy. Our H100 NVL and timing wrapper also differ from their H100 SXM5/C++ setup. Nevertheless, the reversal persists after charging sparse construction: cuSPARSE cannot be treated as an inherently slower CountSketch baseline. We consequently report mixed-precision gains over our own FP32 scatter separately from comparisons with competitive cuSPARSE FP32.

\paragraph{Practical sketch configurations}
Figure~\ref{fig:h200-external} compares FP16 SS with CountSketch scatter and official FlashSketch. Beyond the reasonable fixed four-contribution choice $(\kappa,s,B_r)=(2,2,128)$, we give FlashSketch generous per-case tuning, selecting the fastest configuration meeting FP32 SparseStack's maximum distortion on training seeds, or the minimum-distortion fallback if none qualifies. Choices are frozen before disjoint evaluation; the search guarantees neither optimality nor matched accuracy. As a note, no such tuning was performed for any of the other tested kernels in our test suite, FlashSketch remains the only external evaluation case where we opted to do so.

FP16 SS is fastest among the plotted methods through $d/n=128$, including tuned FlashSketch and CountSketch scatter with fewer contributions. The following experiments test when additional SparseStack contributions justify their cost.

\subsection{Geometry across coherence and sparsity}
\label{subsec:coherence}

\begin{figure*}[!t]
\centering
\includegraphics[width=0.95\textwidth,trim=0pt 5pt 0pt 2pt,clip]{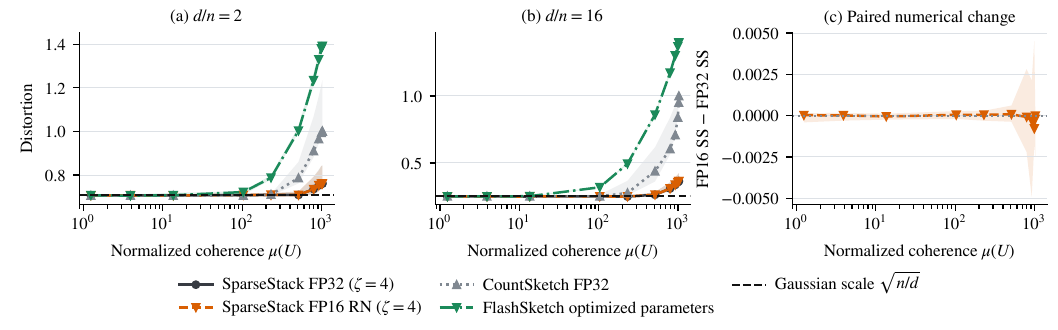}
\caption{Orthogonal-subspace coherence sweep, $m=2^{20}$, $n=1024$, $\zeta=4$. Panel (c) pairs FP16 SS minus FP32 distortion at $d/n=2$. FlashSketch uses frozen optimized parameters. CountSketch FP32 uses the prior full sweep with matching seeded inputs; the remaining curves use the current measurements.}
\label{fig:h200-coherence}
\end{figure*}

We next test how much the complete FP16 kernel changes the geometry retained by an FP32 sketch. The sweep interpolates from a random orthogonal subspace to a coordinate subspace,
\[
 U(\theta)=\operatorname{orth}\bigl((1-\theta)Q_{\rm rand}
                 +\theta[I_n\;0]^T\bigr),\quad 0\le\theta\le1,
\]
where $Q_{\rm rand}$ is an orthogonalized Gaussian matrix. We use eleven $\theta$ values, $m=2^{20}$, and $d/n=2,4,8,16$. Figure~\ref{fig:h200-coherence} reports coherence~\eqref{eq:coherence} and distortion~\eqref{eq:distortion}; $\sqrt{n/d}$ provides a scale reference.

The striking result is how closely FP16 tracks FP32 across this sweep. The largest absolute case-median FP16 SS--FP32 distortion difference is $7.80\times10^{-4}$, and the largest individual difference is 0.00492. Proposition~\ref{prop:distortion} gives a conservative conversion-only bound, with normal-range RMS scale $u\sqrt n=1/64$ here. The measured differences are small even though they also include FP16 and FP32 accumulation errors, which that conversion bound does not control. They demonstrate complete-kernel distortion agreement on these inputs, rather than a sharper theoretical guarantee.

The coherence trend also confirms the expected advantage of spreading each input row across several targets. At $d=2n$, coordinate sketches retain full rank at $\zeta=4,8$, whereas every tested $\zeta=1$ realization loses rank. The handful of $\zeta=2$ realizations we tested also lose rank at $d=2n$; these runs are not plotted in Figure~\ref{fig:h200-coherence}. At the coordinate endpoint and $d/n=16$, FP32 SparseStack has distortion 0.364 versus 1.40 for FlashSketch's tuned fallback. This supports $\zeta=4$ as a useful practical choice, while FlashSketch's block structure makes its behavior sensitive to local neighborhood coherence~\cite[Def.~6.1 and Thm.~6.2]{stanford_flashsketch_paper} explaining the sharp rise in distortion for the highly coherent constructions that we tested.

\subsection{Sketch-and-solve: speed and absolute accuracy}
\label{subsec::empiric::ls}

\begin{figure*}[!t]
\centering
\includegraphics[width=0.95\textwidth,trim=0pt 5pt 0pt 2pt,clip]{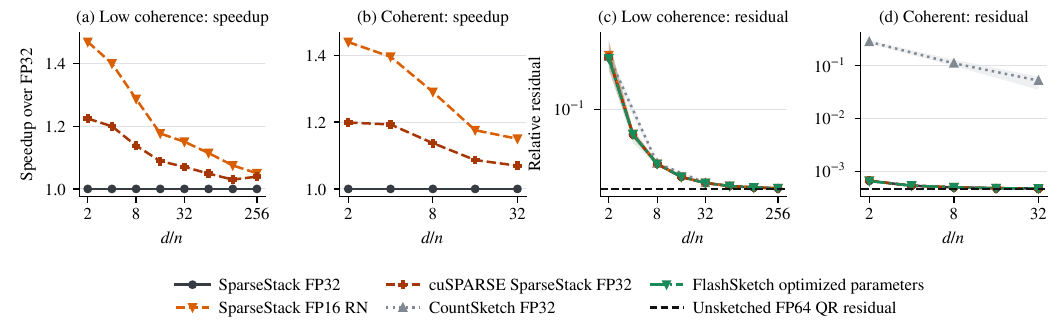}
\caption{Complete LS speedups and original-system residuals. Speedups pair $T_{\rm FP32}/T_{\rm method}$ by seed; values above one indicate a faster solve. Residuals are $\|A\widehat x-b\|/\|b\|$, with unsketched FP64 QR references. Only CountSketch uses a smaller control grid.}
\label{fig:h200-ls}
\end{figure*}

We sketch $A$ in the chosen precision and $b$ in FP32 with the same map, widen FP16 $SA$ exactly, then use FP32 thin QR and a triangular solve without iterative correction. Both main families have $m=2^{21}$, $n=1024$. The low-coherence system has singular values decreasing linearly from 100 to 1 and $b=A\mathbf1+\nu$, $\nu_i\sim\mathcal N(0,0.1^2)$. The coherent system is
\[
 A=\begin{bmatrix}I_n\\T\end{bmatrix},\quad
 T=\frac{10^{-3}G}{\sqrt{m-n}},\quad
 b=\begin{bmatrix}x_0+\nu\\Tx_0\end{bmatrix},
\]
with Gaussian $G,x_0$, head noise $\nu_i\sim\mathcal N(0,0.5^2)$, and a common row permutation. It is well conditioned despite concentrated row energy; the planted $x_0$ is not the noisy system's minimizer.

The faster sketch formation translates into faster complete solves without a comparable change in residual. Figure~\ref{fig:h200-ls} uses our FP32 SparseStack pipeline as the speedup reference. At $d=2n$, FP16 SS gains $1.47\times$ and $1.44\times$ on the low-coherence and coherent systems, respectively, and $1.20\times$ over cuSPARSE on both. Median paired residual increases over FP32 are only $0.34\%$ and $0.35\%$.

Agreement with FP32 sketch-and-solve still leaves sketch approximation error. At $d=2n$, low-coherence coefficient error $\|\widehat x-x_*\|/\|x_*\|$ is 0.444 for FP16 SS and 0.468 for FP32, with $x_*$ the unsketched minimizer. Increasing $d/n$ from two to 32 reduces FP16 SS's residual from 0.1155 to 0.08223, approaching the unsketched FP64 QR reference of 0.08094, but raises wall time from 13.921 to 35.016\,ms. More FP32 factorization work reduces the benefit of faster sketching.

On the coherent system, the expected structural advantage is much more consequential: at $d/n=8$, FP16 SS has residual $4.93\times10^{-4}$ versus 0.112 for CountSketch. SparseStack's additional contributions improve the approximation on this input, and FP16 preserves that improvement. The coherent sweep ends at $d/n=32$.

% \paragraph{Conditioning is a separate boundary}
% An additional $65536\times256$ sweep uses logarithmic spectra at prescribed conditions $1,10^3,10^6$; its coherent head noise is 0.01. At condition $10^6,d/n=8$, FP16 SS completes but has coefficient error 0.994 versus 0.00318 for FP32. Even at condition one, its coherent residual is $3.66\times10^{-4}$ versus $1.08\times10^{-5}$ for FP32. A smaller residual floor exposes rounding, and a completed solve need not recover accurate coefficients.

\subsection{Range boundaries under a common rescaling}
\label{subsec:range}

\begin{figure*}[!t]
\centering
\includegraphics[width=0.95\textwidth,trim=0pt 5pt 0pt 3pt,clip]{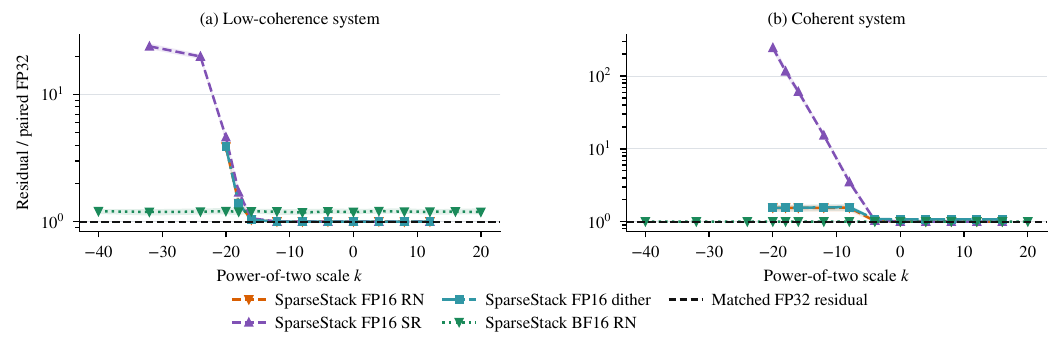}
\caption{Common rescaling at $d/n=2$, $\zeta=4$. FP16 RN uses current measurements; SR, dither, and BF16 are earlier diagnostic controls. Each residual ratio uses its own campaign's paired FP32 reference. Failed settings leave gaps, with coverage retained separately. BF16 is a wider-range reference.}
\label{fig:h200-range}
\end{figure*}

The preceding experiments use input scales at which FP16 largely preserves the FP32 result. A useful operating boundary should also explain when this agreement breaks down. We therefore multiply both $A$ and $b$ by $2^k$: common scaling leaves the mathematical least-squares minimizer and relative residual unchanged, but moves contributions across the FP16 grid. Changes in the computed result then expose sensitivity to numerical representation rather than a different mathematical problem.

The range study uses $65536\times256$ systems, $d/n=2,8$, $\zeta=4$, and fifteen sampled powers from $-40$ to $20$. The low-coherence matrix has singular values spanning 100 to 1 and uses $b=A\mathbf1+\nu$, $\nu_i\sim\mathcal N(0,0.1^2)$. The coherent matrix uses the identity-plus-Gaussian-tail construction above, with tail scale $10^{-3}$, head noise standard deviation 0.5, and a common row permutation; it remains well conditioned. These inputs differ in spectrum, entry scales, and noise placement, so their behavior cannot be attributed to coherence alone. Figure~\ref{fig:h200-range} shows $d/n=2$, with each method compared against its own paired FP32 reference.
% Range generator provenance: the low-coherence RHS/noise recipe follows the user's working confirmation; verify against the generator when available.

\paragraph{Loss of tiny contributions}
At ordinary scales, FP16 SS closely tracks FP32 on both inputs. As the scale decreases, RN erases contributions below half the FP16 subnormal spacing $h=2^{-24}$. The unscaled contribution RMS is $0.1134$ for the low-coherence matrix, but only $1.96\times10^{-6}$ for the coherent tail. The tail therefore reaches the zero-rounding boundary about sixteen powers of two earlier, while the identity head is still representable. Relative dither also loses sufficiently small terms because its perturbation shrinks with their magnitude.

This difference explains the distinct RN trends. Erasing contributions throughout a low-coherence matrix progressively damages the sketch. In the coherent system, however, the retained identity head can keep the reduced system solvable after its weak tail is lost. Once this erasure is complete, the idealized reduced system stops changing after the common scale is undone, consistent with the FP16 SS residual-ratio plateaus of $1.556$ and $1.171$ at $d/n=2$ and eight. The FP32 right-hand-side sketch still contains its tail contribution: the plateau represents a stable altered problem, not recovery of the full-precision result. This mechanism explains the plateau's form without predicting its height.

\paragraph{Stochastic conversion preserves expectation, not accuracy}
SR can preserve the expectation of a tiny contribution through a rare jump to the first nonzero FP16 value. As the input shrinks, such jumps become rarer but larger in original units after the common scale is undone. Corollary~\ref{cor:range-scaling} quantifies this tradeoff: ideal-SR RMS conversion error grows like $2^{-k/2}$. The coherent SR residual slopes are $-0.504$ and $-0.506$ over $k=-20$ through $-8$, consistent with that mechanism. Transferring the exponent to residuals requires approximately linear propagation through a stable solve with perturbations above the residual floor; the implemented finite-bit SR and accumulation are not isolated by this experiment.

At $k=-20,d/n=2$, SR gives relative residual $0.165$, about $244\times$ its paired FP32 value, while FP16 SS gives $0.00103$. The SR sketch still has full rank and condition about $6.42$, so the poor result is not explained by rank loss. The small coherent residual floor amplifies the ratio, but the absolute residual also shows a substantial loss of accuracy. Retaining tiny terms in expectation can therefore be less useful than RN's deterministic erasure of them on this input.

\paragraph{Representation limits and failed solves}
At sufficiently small scales, even the coherent head is erased: FP16 SS fails at $k=-40,-32,-24$ on both constructions. In particular, at $k=-24$ the identity contribution lies exactly at $h/2$ and nearest-even conversion returns zero. At the other extreme, large contributions or running sums can exceed finite FP16 range. FP16 SS fails at $k=16,20$ on the low-coherence system and at $20$ on the coherent one; the coherent head contribution is still finite at $k=16$, with value $32768$.

SR completes some additional tiny low-coherence settings but with poor residuals. BF16 remains finite throughout the sampled range, yet its ordinary-scale low-coherence residual is about $20\%$ larger than FP32 at $d/n=2$. Wider range thus avoids one failure mechanism without removing the accuracy cost of reduced precision. Useful FP16 operation requires both contributions within its usable range and acceptable complete-sketch error; successful completion alone establishes neither.

\section{Discussion and Conclusions}
\label{sec::discussion}

Our results support reduced precision as a practical algorithmic choice in RandNLA. Packed FP16 updates accelerate SparseStack while closely preserving FP32 geometry and least-squares residuals at ordinary input scales. Under stated assumptions, random signs cancel conversion-error cross terms, giving RN no greater expected squared conversion error than ideal SR and bounding conversion-induced distortion. Accumulation remains outside this guarantee. The range experiments show that retaining tiny terms in expectation can increase variance, while RN can erase information; stochastic conversion need not improve complete-solve accuracy. RN also avoids randomized-conversion overhead. These findings motivate checking input range and application accuracy before reducing precision.

This opens several promising avenues for future research:
\begin{itemize}[leftmargin=*,itemsep=2pt,topsep=3pt,parsep=0pt]
    \item \textbf{Adaptive sketching cascades:} Use FP8/FP4 sketches as fast filters or coarse preconditioners, refining when residual or distortion tests indicate insufficient accuracy.
    % \item \textbf{Low-precision dense projections:} Extend the rounding and sign-cancellation analysis to dense Gaussian sketches on Tensor Cores.
    \item \textbf{End-to-end multi-precision solvers:} Couple low-precision downstream QR or Cholesky with iterative refinement; our current solve remains FP32.
    \item \textbf{Dynamic exponent scaling:} Track blockwise exponents efficiently to prevent underflow and preserve dynamic range without manual global normalization.
\end{itemize}

% FP16 SS improves every complete-sketch case median in the $\zeta=1,2,4,8$ timing grid with the same arithmetic pipeline. At four contributions it beats custom FP32 and complete cuSPARSE FP32; cuSPARSE remains the stronger CountSketch baseline. FP32 QR limits complete-LS gains.

% Random signs cancel conversion cross terms, giving RN no larger expected squared conversion error than ideal SR. The distortion bound transfers this result to conversion-induced geometric error. Both bounds are conservative; correlated accumulation errors remain outside this guarantee. The small complete-kernel distortion differences and coherent-input advantage are empirical findings, and do not establish coefficient accuracy for arbitrary conditioning.

% Common rescaling can erase matrix information retained by the FP32 RHS or increase stochastic-conversion variance despite an unchanged mathematical problem. Wider range avoids failure without guaranteeing accuracy. Ordinary RN is practical when the complete sketch error meets the application tolerance, with conditioning, distribution, and input scale checked.

\bibliographystyle{IEEEtran}
\bibliography{references}

@String{Computing = "Computing" }

@String{Computer = "{IEEE} Computer" }

@misc{randlapack_book,
      title={Randomized Numerical Linear Algebra : A Perspective on the Field With an Eye to Software}, 
      author={Riley Murray and James Demmel and Michael W. Mahoney and N. Benjamin Erichson and Maksim Melnichenko and Osman Asif Malik and Laura Grigori and Piotr Luszczek and Michał Dereziński and Miles E. Lopes and Tianyu Liang and Hengrui Luo and Jack Dongarra},
      year={2023},
      eprint={2302.11474},
      archivePrefix={arXiv},
      primaryClass={math.NA},
      url={https://arxiv.org/abs/2302.11474}, 
}

@INPROCEEDINGS{OG_sketching_sarlos,
  author={Sarlos, Tamas},
  booktitle={2006 47th Annual IEEE Symposium on Foundations of Computer Science (FOCS'06)}, 
  title={Improved Approximation Algorithms for Large Matrices via Random Projections}, 
  year={2006},
  volume={},
  number={},
  pages={143-152},
  doi={10.1109/FOCS.2006.37},
  url={https://doi.org/10.1109/FOCS.2006.37}}

@article{woodruff_sketching,
  author       = {David P. Woodruff},
  title        = {Sketching as a Tool for Numerical Linear Algebra},
  journal      = {CoRR},
  volume       = {abs/1411.4357},
  year         = {2014},
  url          = {http://arxiv.org/abs/1411.4357},
  eprinttype   = {arXiv},
  eprint       = {1411.4357},
  bibsource    = {dblp computer science bibliography, https://dblp.org}
}

@book{Chen_online_book,
    title = {Randomized Numerical Linear Algebra with Examples},
    author = {Tyler Chen},
    year = {2025},
    version = {prerelease},
    url = {https://research.chen.pw/RandNLA},
}

@misc{chennakod_sparse_stack_theory,
      title={Optimal Subspace Embeddings: Resolving Nelson-Nguyen Conjecture Up to Sub-Polylogarithmic Factors}, 
      author={Shabarish Chenakkod and Michał Dereziński and Xiaoyu Dong},
      year={2025},
      eprint={2508.14234},
      archivePrefix={arXiv},
      primaryClass={cs.DS},
      url={https://arxiv.org/abs/2508.14234}, 
}

@misc{chen_GPU_sketching_paper,
      title={GPU-Parallelizable Randomized Sketch-and-Precondition for Linear Regression using Sparse Sign Sketches}, 
      author={Tyler Chen and Pradeep Niroula and Archan Ray and Pragna Subrahmanya and Marco Pistoia and Niraj Kumar},
      year={2025},
      eprint={2506.03070},
      archivePrefix={arXiv},
      primaryClass={cs.DS},
      url={https://arxiv.org/abs/2506.03070}, 
}

@misc{higgins_GPU_sketching_paper,
      title={A High Performance GPU CountSketch Implementation and Its Application to Multisketching and Least Squares Problems}, 
      author={Andrew J. Higgins and Erik G. Boman and Ichitaro Yamazaki},
      year={2025},
      eprint={2508.14209},
      archivePrefix={arXiv},
      primaryClass={math.NA},
      url={https://arxiv.org/abs/2508.14209}, 
}

@misc{stanford_flashsketch_paper,
      title={FlashSketch: Sketch-Kernel Co-Design for Fast Sparse Sketching on GPUs}, 
      author={Rajat Vadiraj Dwaraknath and Sungyoon Kim and Mert Pilanci},
      year={2026},
      eprint={2602.06071},
      archivePrefix={arXiv},
      primaryClass={cs.DC},
      url={https://arxiv.org/abs/2602.06071}, 
}

@misc{camano_structured_sketching_tropp,
      title={Faster Linear Algebra Algorithms with Structured Random Matrices}, 
      author={Chris Camaño and Ethan N. Epperly and Raphael A. Meyer and Joel A. Tropp},
      year={2025},
      eprint={2508.21189},
      archivePrefix={arXiv},
      primaryClass={cs.DS},
      url={https://arxiv.org/abs/2508.21189}, 
}

@article{charikar_countsketch_init_paper,
title = {Finding frequent items in data streams},
journal = {Theoretical Computer Science},
volume = {312},
number = {1},
pages = {3-15},
year = {2004},
note = {Automata, Languages and Programming},
issn = {0304-3975},
doi = {https://doi.org/10.1016/S0304-3975(03)00400-6},
url = {https://www.sciencedirect.com/science/article/pii/S0304397503004006},
author = {Moses Charikar and Kevin Chen and Martin Farach-Colton}
}

@misc{martinsson_tropp_randnla_review,
      title={Randomized Numerical Linear Algebra: Foundations \& Algorithms}, 
      author={Per-Gunnar Martinsson and Joel Tropp},
      year={2021},
      eprint={2002.01387},
      archivePrefix={arXiv},
      primaryClass={math.NA},
      url={https://arxiv.org/abs/2002.01387}, 
}

@misc{Higham_stochastic_blog,
  author       = {Higham, Nicholas J.},
  title        = {What Is Stochastic Rounding?},
  howpublished = {Nicholas Higham (blog)},
  year         = {2020},
  month        = {July},
  day          = {7},
  url          = {https://nhigham.com/2020/07/07/what-is-stochastic-rounding/}
}

@article{connolly_higham_stochastic,
author = {Connolly, Michael P. and Higham, Nicholas J. and Mary, Theo},
title = {Stochastic Rounding and Its Probabilistic Backward Error Analysis},
journal = {SIAM Journal on Scientific Computing},
volume = {43},
number = {1},
pages = {A566-A585},
year = {2021},
doi = {10.1137/20M1334796},

URL = { 
    
        https://doi.org/10.1137/20M1334796
    
    

},
eprint = { 
    
        https://doi.org/10.1137/20M1334796
    
    

}
}

@misc{dithering_stanford_pilanci,
  author       = {{Mert Pilanci}},
  title        = {Lecture Slides: Dithering},
  howpublished = {EE269: Signal Processing for Machine Learning},
  year         = {2026},
  note         = {Available at: \url{https://web.stanford.edu/class/ee269/Lecture_dithering.pdf}},
  url          = {https://web.stanford.edu/class/ee269/Lecture_dithering.pdf}
}

@misc{grass_paper,
      title={GraSS: Scalable Data Attribution with Gradient Sparsification and Sparse Projection}, 
      author={Pingbang Hu and Joseph Melkonian and Weijing Tang and Han Zhao and Jiaqi W. Ma},
      year={2025},
      eprint={2505.18976},
      archivePrefix={arXiv},
      primaryClass={cs.LG},
      url={https://arxiv.org/abs/2505.18976}, 
}

@article{haidar_mixed_precision_ir,
    author = {Haidar, Azzam and Bayraktar, Harun and Tomov, Stanimire and Dongarra, Jack and Higham, Nicholas J.},
    title = {Mixed-precision iterative refinement using tensor cores on GPUs to accelerate solution of linear systems},
    journal = {Proceedings of the Royal Society A: Mathematical, Physical and Engineering Sciences},
    volume = {476},
    number = {2243},
    pages = {20200110},
    year = {2020},
    month = {11},
    issn = {1364-5021},
    doi = {10.1098/rspa.2020.0110},
    url = {https://doi.org/10.1098/rspa.2020.0110},
    eprint = {https://royalsocietypublishing.org/rspa/article-pdf/doi/10.1098/rspa.2020.0110/639311/rspa.2020.0110.pdf},
}

@article{carson_higham_three_precision,
author = {Carson, Erin and Higham, Nicholas J.},
title = {Accelerating the Solution of Linear Systems by Iterative Refinement in Three Precisions},
journal = {SIAM Journal on Scientific Computing},
volume = {40},
number = {2},
pages = {A817-A847},
year = {2018},
doi = {10.1137/17M1140819},

URL = { 
    
        https://doi.org/10.1137/17M1140819
    
    

},
eprint = { 
    
        https://doi.org/10.1137/17M1140819
    
    

}
}

@article{higham_squeezing_half_precision,
author = {Higham, Nicholas J. and Pranesh, Srikara and Zounon, Mawussi},
title = {Squeezing a Matrix into Half Precision, with an Application to Solving Linear Systems},
journal = {SIAM Journal on Scientific Computing},
volume = {41},
number = {4},
pages = {A2536-A2551},
year = {2019},
doi = {10.1137/18M1229511},

URL = { 
    
        https://doi.org/10.1137/18M1229511
    
    

},
eprint = { 
    
        https://doi.org/10.1137/18M1229511
    
    

}
}

@article{hpl_mxp_benchmark,
   title={HPL-MxP benchmark: Mixed-precision algorithms, iterative refinement, and scalable data generation},
   volume={40},
   ISSN={1741-2846},
   url={http://dx.doi.org/10.1177/10943420251382476},
   DOI={10.1177/10943420251382476},
   number={1},
   journal={The International Journal of High Performance Computing Applications},
   publisher={SAGE Publications},
   author={Dongarra, Jack and Luszczek, Piotr},
   year={2025},
   month=Sept, pages={52–62} }

@article{wannamaker_dither_survey,
author = {Lipshitz, Stanley and Wannamaker, Rob and Vanderkooy, John},
year = {1992},
month = {05},
pages = {355-374},
title = {Quantization and Dither: A Theoretical Survey},
volume = {40},
journal = {Journal of the Audio Engineering Society},
url = {https://aes2.org/publications/elibrary-page/?id=7047}
}

@phdthesis{wannamaker_dithering_thesis,
  author       = {Wannamaker, Robert Alexander},
  title        = {The Theory of Dithered Quantization},
  school       = {University of Waterloo},
  year         = {1997},
  address      = {Waterloo, Ontario, Canada},
  type         = {Doctor of Philosophy thesis},
  url          = {https://uwspace.uwaterloo.ca/items/3e8336cc-a6ea-43c9-bdbc-7256bfdbcca8}
}

@inbook{sparse_sign_ose_bound,
author = {Michael B. Cohen},
title = {Nearly Tight Oblivious Subspace Embeddings by Trace Inequalities},
booktitle = {Proceedings of the 2016 Annual ACM-SIAM Symposium on Discrete Algorithms (SODA)},
chapter = {},
pages = {278-287},
doi = {10.1137/1.9781611974331.ch21},
URL = {https://epubs.siam.org/doi/abs/10.1137/1.9781611974331.ch21},
eprint = {https://epubs.siam.org/doi/pdf/10.1137/1.9781611974331.ch21}
}

@misc{carson_dauzickaite_mxp_sketching_ls,
      title={Mixed Precision Sketching for Least-Squares Problems and its Application in {GMRES}-Based Iterative Refinement},
      author={Erin Carson and Ieva Dau\v{z}ickait\.{e}},
      year={2024},
      eprint={2410.06319},
      archivePrefix={arXiv},
      primaryClass={math.NA},
      url={https://arxiv.org/abs/2410.06319},
}

@misc{carson_dauzickaite_mxp_nystrom,
      title={Single-pass Nystr\"{o}m approximation in mixed precision}, 
      author={Erin Carson and Ieva Dau\v{z}ickait\.{e}},
      year={2023},
      eprint={2205.13355},
      archivePrefix={arXiv},
      primaryClass={math.NA},
      url={https://arxiv.org/abs/2205.13355}, 
}

@inproceedings{ootomo_yokota_mxp_sketching,
  author = {Hiroyuki Ootomo and Rio Yokota},
  title = {Mixed-Precision Random Projection for {RandNLA} on {Tensor Cores}},
  booktitle = {Proceedings of the Platform for Advanced Scientific Computing Conference},
  series = {PASC '23},
  year = {2023},
  publisher = {Association for Computing Machinery},
  doi = {10.1145/3592979.3593413},
  url = {https://doi.org/10.1145/3592979.3593413}
}

@manual{nvidia_ptx_isa,
  author = {{NVIDIA Corporation}},
  title = {{PTX ISA}, Version 9.4: Parallel Synchronization and Communication Instructions, {atom}},
  year = {2026},
  url = {https://docs.nvidia.com/cuda/parallel-thread-execution/index.html#parallel-synchronization-and-communication-instructions-atom}
}

@article{JL_lemma_og,
  title={Extensions of Lipschitz mappings into Hilbert space},
  author={William B. Johnson and Joram Lindenstrauss},
  journal={Contemporary mathematics},
  year={1984},
  volume={26},
  pages={189-206},
  url={https://api.semanticscholar.org/CorpusID:117819162}
}
\end{document}